\documentclass[aps,prl,preprint,superscriptaddress]{revtex4-1}

\usepackage{amsmath}
\usepackage{amssymb}
\usepackage{graphicx}
\usepackage{notes2bib}
\usepackage{xcolor}

\usepackage[colorlinks = true,
linkcolor = blue,
urlcolor  = blue,
citecolor = blue,
anchorcolor = blue,
breaklinks = true]{hyperref}

\bibnotesetup{
note-name = ,
use-sort-key = false,
placement = mixed
}

\begin{document}

%\preprint{}

\title{Unified view of nonlinear wave structures associated with whistler-mode chorus}


\author{Xin An}
\email[]{xinan@atmos.ucla.edu}
%\thanks{}
\affiliation{Department of Atmospheric and Oceanic Sciences, University of California, Los Angeles, California, USA.}

\author{Jinxing Li}
\affiliation{Department of Atmospheric and Oceanic Sciences, University of California, Los Angeles, California, USA.}

\author{Jacob Bortnik}
\affiliation{Department of Atmospheric and Oceanic Sciences, University of California, Los Angeles, California, USA.}

\author{Viktor Decyk}
\affiliation{Department of Physics and Astronomy, University of California, Los Angeles, California, USA.}

\author{Craig Kletzing}
\affiliation{Department of Physics and Astronomy, University of Iowa, Iowa City, Iowa, USA.}

\author{George Hospodarsky}
\affiliation{Department of Physics and Astronomy, University of Iowa, Iowa City, Iowa, USA.}


\date{\today}

\begin{abstract}
A range of nonlinear wave structures, including Langmuir waves, unipolar electric fields and bipolar electric fields, are often observed in association with whistler-mode chorus waves in the near-Earth space. We demonstrate that the three seemingly different nonlinear wave structures originate from the same nonlinear electron trapping process by whistler-mode chorus waves. The ratio of the Landau resonant velocity to the electron thermal velocity controls the type of nonlinear wave structures that will be generated.
\end{abstract}

\pacs{}
%\keywords{}

\maketitle

%\section{}
%% Put \label in argument of \section for cross-referencing
%%\section{\label{}}
%\subsection{}
%\subsubsection{}

Whistler-mode chorus \cite{tsurutani1974postmidnight, coroniti1980detection, hospodarsky2008observations} is a coherent electromagnetic emission found widely in the near-space region of the Earth and other magnetized planets. Chrous waves are the Earth's own ``cyclotron accelerator'' that accelerates the radiation belt electrons \cite{horne2005timescale, thorne2013rapid}. They can also scatter the energetic electrons out of their trapped orbit and light up the pulsating aurora in the upper atmosphere \cite{nishimura2010identifying}. Nonlinear wave structures, for example, Langmuir waves \cite{reinleitner1982chorus, reinleitner1984chorus, li2017chorus, li2018local}, unipolar electric fields \cite{kellogg2010electron, mozer2014direct, gao2016generation, vasko2018electrostatic, agapitov2018nonlinear, malaspina2018census} and bipolar electric fields \cite{wilder2016observations, malaspina2018census}, are often observed in association with chorus waves. These nonlinear wave structures are considered to be important since they have the potential for significant particle scattering and acceleration \cite{mozer2014direct, artemyev2014thermal, mozer2015time, vasko2017diffusive}. Despite several past attempts \cite{reinleitner1983electrostatic, bujarbarua1985plasma, drake2015development, gao2017generation, vasko2018electrostatic, agapitov2018nonlinear} to explain the generation of such nonlinear wave structures and their relation to chorus, their linkage is not yet understood, and direct measurements of electron phase space structures responsible for these nonlinear wave structures have been difficult to obtain.

In this Letter we demonstrate the link between several different nonlinear wave structures and whistler-mode chorus, by observing the associated electron phase space structures using computer simulations. When the tail of the electron distribution is trapped by chorus, trapped electrons form a spatially modulated bump-on-tail distribution and excite Langmuir waves. When the thermal electrons are trapped by chorus, they form phase space holes and hence produce bipolar electric fields. Between these two regimes, trapped electrons generate nonlinear electron acoustic waves, which in turn disrupt the trapped electrons and accumulates them in a limited spatial region, leading to the unipolar electric fields. This study connects a variety of seemingly unrelated nonlinear field structures and provides a simple, integrated picture of the microscopic interactions between whistler waves and electrons.

\begin{figure*}[tphb]
\centering
\includegraphics[width=6.75in]{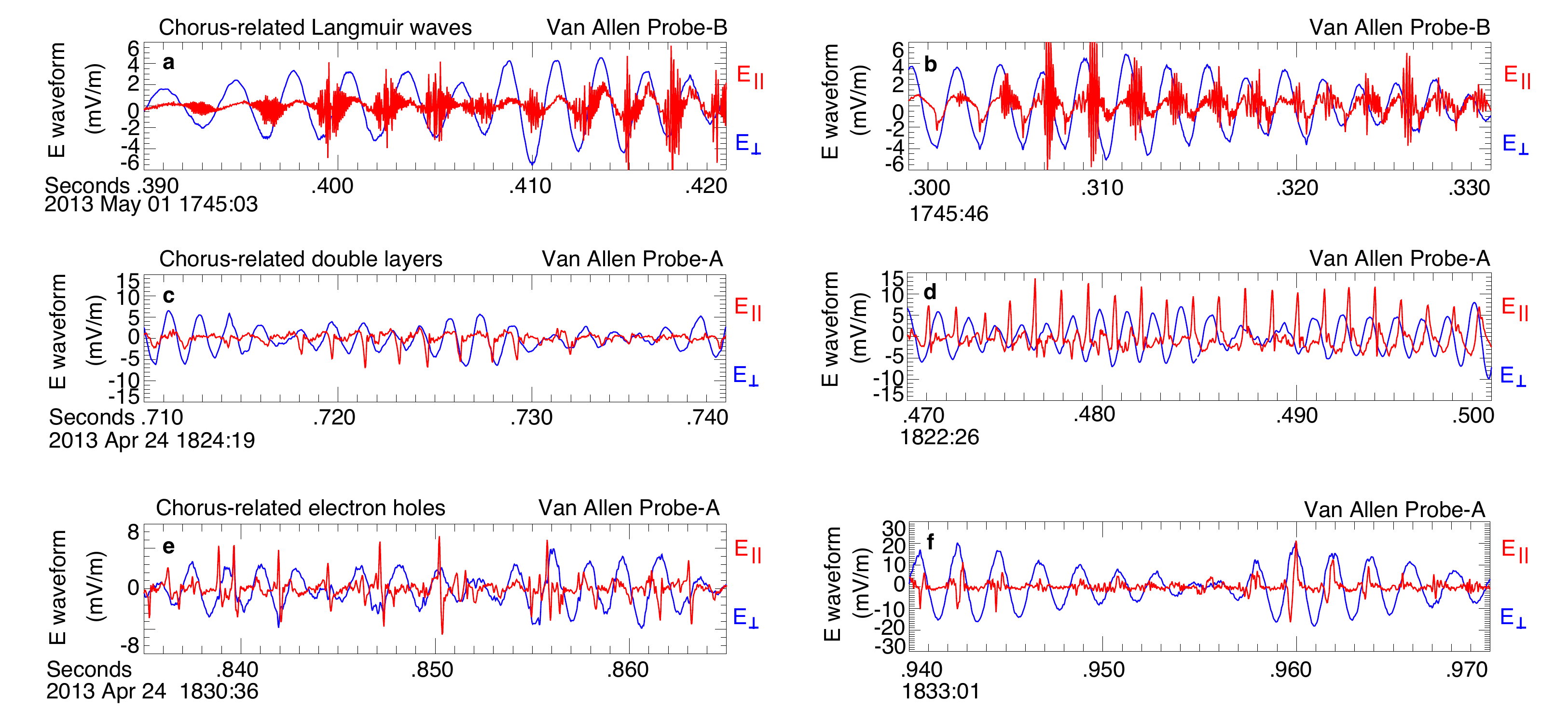}
\caption{\label{fig1}Typical nonlinear wave structures associated with whistler-mode chorus waves from spacecraft observations. (a)-(f) The parallel (red) and perpendicular (blue) electric waveforms measured by the EMFISIS instrument aboard Van Allen Probes. (a), (c), (e), example with chorus propagation parallel to the background magnetic field. (b), (d), (f), example with chorus propagation anti-parallel to the background magnetic field. (a)-(b) chorus-related Langmuir waves. (c)-(d) chorus-related unipolar electric fields. (e)-(f) chorus-related bipolar electric fields.}
\end{figure*}

The three basic types of nonlinear wave structures are illustrated using data from the Electric and Magnetic Field Instrument Suite and Integrated Science (EMFISIS) \cite{kletzing2013electric} on board NASA's Van Allen Probes. High-frequency Langmuir waves are seen to occur primarily near the negative phase of the whistler parallel electric field (i.e., parallel with respect to background magnetic field) and shown in Fig.~\ref{fig1}(a). Langmuir waves are a class of electrostatic plasma waves, naturally found in the Earth's near-space environment at frequencies near the electron plasma frequency \cite{reinleitner1982chorus, li2017chorus} ($\omega_{pe}$, corresponding to the electrostatic oscillation frequency of electrons in response to a small charge separation). In the other examples, the parallel electric field of the chorus is highly distorted and appears as either a unipolar electric field [Fig.~\ref{fig1}(c)] or a bipolar electric field structure [Fig.~\ref{fig1}(e)]. The unipolar electric field is also called a `double layer' because it resembles a net potential drop from a layer of net positive charges to an adjacent layer of net negative charges, whereas the bipolar electric field is also referred to as an `electron hole' since it resembles the field created by a collection of positive charges. As the propagation direction of whistler is reversed (See Supplemental Material \bibnote{See Supplemental Material for the reversal of the Poynting flux direction} for the propagation direction), the excitation location of Langmuir waves changes to occur primarily near the positive phase of the whistler parallel electric field [Fig.~\ref{fig1}b]. The polarity of unipolar electric fields and bipolar electric fields also change to the opposite sense [Fig.~\ref{fig1}(d), Fig.~\ref{fig1}(f)]. This reversal in the association of whistler wave phase and electric field polarity indicates that such nonlinear wave structures are closely associated and possibly driven by whistler-mode chorus waves. The nontrivial, observable features of the nonlinear wave structures here are distinct from those directly driven by electron beams \cite[e.g.,][]{omura1996electron}.

% figure 1

\begin{figure}[tphb]
\centering
\includegraphics[width=3.375in]{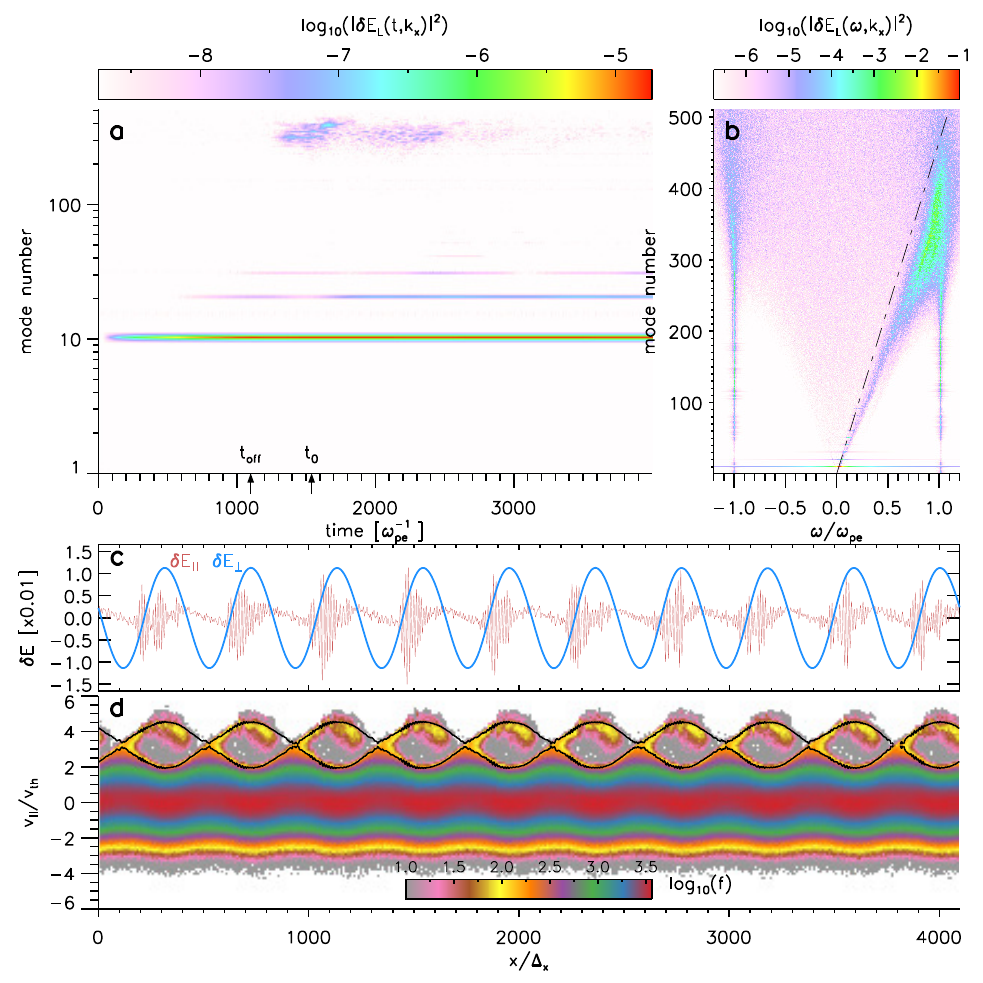}
\caption{\label{fig2}Langmuir waves driven by whistler-mode chorus in Simulation 1 with $v_r / v_{th} = 3.2$. (a) The temporal evolution of the wave number spectrum of the longitudinal electric field $\delta E_L$ (along the direction of wave propagation). The horizontal axis is shown in units of time normalized to the plasma frequency $\omega_{pe}$. The pump field is turned off at $t_{\text{off}} = 1100\, \omega_{pe}^{-1}$. The mode number of whistler-mode chorus is $10$, which represents the number of wave lengths in the system. Two harmonics of whistler occur consecutively in time at mode numbers $20$ and $30$. Langmuir waves are located mainly in the mode number range $300$-$400$. (b) The dispersion diagram, obtained by Fourier-transforming the wave number spectrum from the time domain to the frequency domain. The whistler frequency is $0.0215\, \omega_{pe}$. Beam-mode Langmuir waves are found just below $\omega_{pe}$. The reciprocal of the slope of the dash-dotted line represents the phase velocity of whistler, which is the same as that of the whistler harmonics, but is smaller than that of Langmuir waves. (c) The waveforms of the electric fields parallel (red) and perpendicular (blue) to $B_0$ at $t_0 = 1540\, \omega_{pe}^{-1}$. The horizontal axis is the spatial coordinate $x$ normalized by grid size $\Delta_x$. Langmuir waves are seen in $\delta E_\parallel$. (d) The phase space portrait at $t_0 = 1540\, \omega_{pe}^{-1}$. The magnitude of phase space density $f$ is coded in color. The separatrix (black line) encircles the resonant island, dividing trapped and untrapped particles. Trapped particles are bounded in phase in the wave frame of the propagating whistler waves whereas untrapped particles are able to freely oscillate in phase.}
\end{figure}

To gain insight into the generation process of the nonlinear wave structures, we performed a series of 1D spatial, 3D velocity Particle-In-Cell (PIC) simulations \cite{decyk2007upic, kaufman1971darwin, nielson1976particle, busnardo1977self} which are able to capture complex nonlinear interactions between the whistler waves and any electrons that are potentially trapped by the wave's electromagnetic fields (See Supplemental Material \bibnote{See Supplemental Material for details of the simulation setup} for details of simulation setup). We set up the whistler wave field by driving the plasma with an external pump field for a prescribed time interval. After the pump field is turned off, the electromagnetic field of whistler wave continues to propagate, and is self-consistently supported by the electron distribution. In all the simulations, whistler waves reach an amplitude of $\lesssim 0.1 B_0$ and propagate at an angle of $30^\circ$ with respect to $B_0$. Here $B_0$ is the background magnetic field. The whistler electric field parallel to $B_0$ can be in Landau resonance with electrons, where the electron velocity matches the whistler phase velocity parallel to $B_0$. Therefore the whistler parallel electric field can trap the resonant electrons in its potential well. We denote the Landau resonant velocity as $v_r$. By varying the ratio of the Landau resonant velocity $v_r$ to the initial electron thermal velocity $v_{th}$, we observe three typical types of nonlinear wave structures [Fig.~\ref{fig2}(c), Fig.~\ref{fig3-distribution}(c) and Fig.~\ref{fig4}(c)] that agree remarkably well with the types of nonlinear structures that spacecraft observations show [Fig.~\ref{fig1}(a), Fig.~\ref{fig1}(c), Fig.~\ref{fig1}(e)]. The three regimes of nonlinear wave generation are described below.

In our simulation, Langmuir waves are excited when whistler waves resonate with electrons at the tail of the distribution. An example with $v_r / v_{th} = 3.2$ is shown in Fig.~\ref{fig2}. The phase space of electrons is displayed as a function of the wave propagation direction $x$ and the electron parallel velocity $v_{\parallel}$. Electrons around the resonant velocity ($v_r / v_{th} = 3.2$) get trapped in the resonant island by whistler waves. They are accelerated in the portion of $\delta E_{\parallel} < 0$, stream near the separatrix of the resonant island and form a spatially modulated bump-on-tail distribution [See Fig.~\ref{fig2}(d) and Supplementary Video 1]. Therefore the localized bump-on-tail distribution excites beam-mode Langmuir waves \cite{o1968transition} primarily near the negative phase of $\delta E_{\parallel}$ [Fig.~\ref{fig2}(b)-(c)]. Eventually, the excited Langmuir waves tend to diffuse the bump-on-tail distribution and gradually flatten the distribution in the resonant island. The modest spatial bunching of the trapped electrons in the resonant island also leads to weak harmonics [Fig.~\ref{fig2}(a)] of the fundamental wave number of the whistler waves. To test our hypothesis further, we confirmed that Langmuir waves are excited near the positive phase of $\delta E_{\parallel}$ by reversing the propagation direction of whistler waves (not shown), consistent with observations.

% figure 2

\begin{figure}[tphb]
\centering
\includegraphics[width=3.375in]{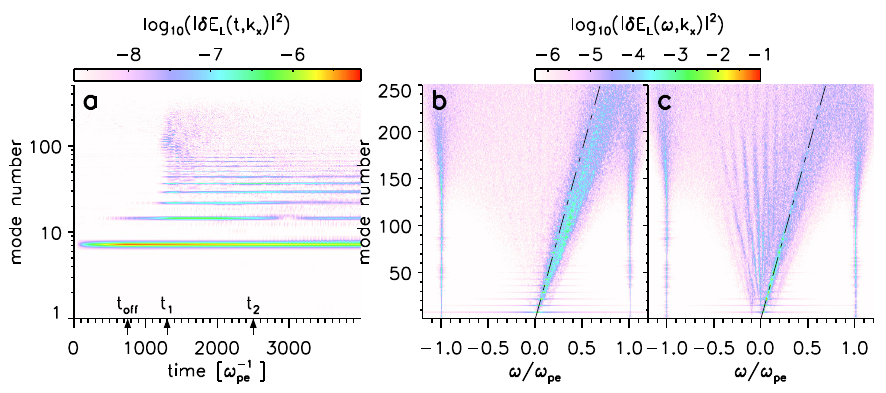}
\caption{\label{fig3-spectra}The wave number spectrum and dispersion diagram in Simulation 2 with $v_r / v_{th} = 2.1$. (a) The temporal evolution of the wave number spectrum of the longitudinal electric field. The pump field is turned off at $t_{\text{off}} = 750\, \omega_{pe}^{-1}$. The mode number of whistler wave is $7$. Electron acoustic waves start to be excited in the interval $1000 < t\omega_{pe} < 2000$. (b) The dispersion diagram for $0 < t \omega_{pe} \leqslant 2000$. The phase velocity of electron acoustic waves is slightly larger than that of the whistler waves (dash-dotted line). (c) The dispersion diagram for $2000 < t \omega_{pe} \leqslant 4000$, showing that all the whistler harmonics propagate at the same phase speed as fundamental whistler wave itself. Beam-mode waves with phase velocities smaller than that of whistler wave propagate in both forward and backward directions.}
\end{figure}

\begin{figure}[tphb]
\centering
\includegraphics[width=3.375in]{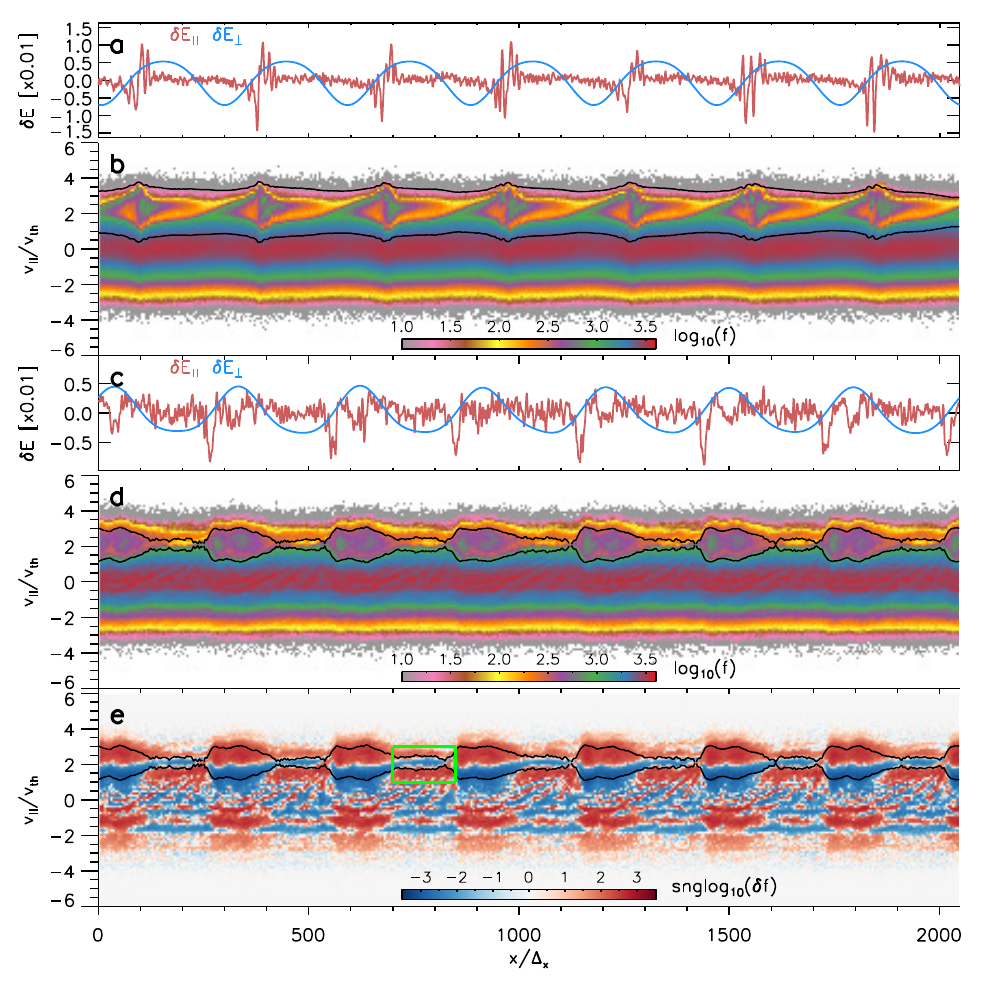}
\caption{\label{fig3-distribution}The distortion of whistler wave electric field into unipolar electric field in Simulation 2 with $v_r / v_{th} = 2.1$. (a) The waveforms of the electric fields parallel (red) and perpendicular (blue) to $B_0$ at $t_1 = 1300\, \omega_{pe}^{-1}$. Electron acoustic waves are seen in $\delta E_\parallel$. (b) The phase space portrait at  $t_1 = 1300\, \omega_{pe}^{-1}$. A spatially modulated beam is formed within the separatrix (black line). (c) The waveforms of the electric fields parallel (red) and perpendicular (blue) to $B_0$ at $t_2 = 2500\, \omega_{pe}^{-1}$. The whistler electric field is distorted into unipolar electric field shown in $\delta E_\parallel$. (d)-(e) The phase space portrait at $t_2 = 2500\, \omega_{pe}^{-1}$, color coded by phase space density (panel d) and perturbed phase space density (panel e). The signed logarithm of the perturbed phase space density $\operatorname{snglog}_{10}(\delta f)$ is defined as $\operatorname{sng}(\delta f) \cdot \operatorname{log}_{10}(\lvert \delta f \rvert)$. The electrons are accumulated in phases where the separatrix (black line) is narrowed. An example of electron accumulation is indicated by the green box. Several small beams are seen in the velocity range $-2 < v_{\parallel}/v_{th} < 1$.}
\end{figure}

\begin{figure}[tphb]
\centering
\includegraphics[width=3.375in]{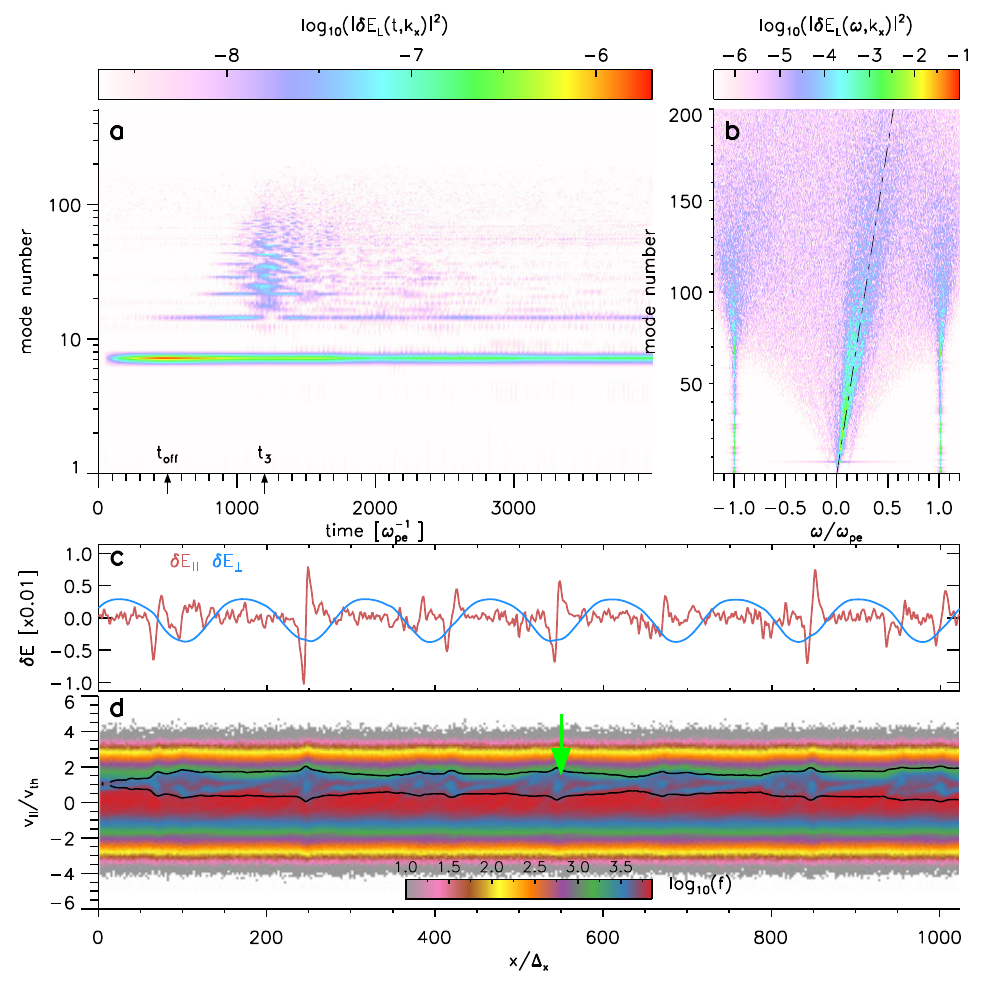}
\caption{\label{fig4}Bipolar electric field structures driven by whistler-mode chorus waves in Simulation 3 with $v_r / v_{th} = 1.0$. (a) The temporal evolution of the wave number spectrum of the longitudinal electric field. The pump field is turned off at $t_{\text{off}} = 500\, \omega_{pe}^{-1}$. The mode number of whistler wave is $7$. The transient broadband spectrum with mode numbers ranging between $20$ and $100$ corresponds to the bipolar field structures. (b) The dispersion diagram showing that the phase velocity of bipolar structures is around that of whistler (dash-dotted line). (c) The waveforms of the electric fields parallel (red) and perpendicular (blue) to $B_0$ at $t_3 = 1200\, \omega_{pe}^{-1}$. Bipolar electric fields are seen in $\delta E_\parallel$. (d) The phase space portrait at $t_3 = 1200\, \omega_{pe}^{-1}$. Each phase space hole is a region of electron deficit inside the separatrix (black line), generating a bipolar electric field structure. An example of phase space hole is indicated by the green arrow.}
\end{figure}

As the whistler waves begin to resonate with electrons closer to the bulk of the distribution, the unipolar electric field structure starts to become more prevalent. An example with $v_r / v_{th} = 2.1$ is shown in Fig.~\ref{fig3-spectra} and Fig.~\ref{fig3-distribution}. Electrons trapped by whistler waves form a spatially modulated beam [Fig.~\ref{fig3-distribution}(b)] in the region of high phase space density instead of in the tail. The electron beam also generates electrostatic beam-mode waves [Fig.~\ref{fig3-spectra}(b), Fig.~\ref{fig3-distribution}(a)], oscillating at smaller wave frequencies compared to Langmuir waves. These beam-mode waves are identified as nonlinear electron acoustic mode waves \cite{holloway1991undamped, valentini2006excitation, anderegg2009electron}. They have their phase velocity located within the resonant island and therefore can survive undamped on the plateau of the distribution in this region. The beam-generated electron acoustic waves disrupt the separatrix of the original resonant island and transport the originally trapped electrons such that they accumulate in a limited range of phase outside the newly formed separatrix [See Fig.~\ref{fig3-distribution}(d)-(e) and Supplemental Video 2]. The accumulation of electrons leads to a pronounced unipolar electric field in the spatial domain [Fig.~\ref{fig3-distribution}(c)]. This unipolar electric field is directed from the phase of the adjacent resonant island to the phase of electron accumulation. The spatial scale of the unipolar structure is about a few tens of Debye lengths. The harmonics of whistler waves in the wave spectrum [Fig.~\ref{fig3-spectra}(a)] are a manifestation of the unipolar structure, which have the same phase velocity as that of the fundamental whistler waves [Fig.~\ref{fig3-spectra}(c)]. Contrary to the transient beam instability and beam-mode waves, the phase space structure associated with the unipolar electric field is long-lived. It is interesting to note that the bulk of the distribution at $-2 < v_{\parallel}/v_{th} < 1$ is structured to form a few small beams [Fig.~\ref{fig3-distribution}(d)-(e)] after the major beam instability. These beams also generate beam-mode waves, propagating in small phase velocities in both forward and backward directions [Fig.~\ref{fig3-spectra}(c)]. It is also worthwhile to note that the net potential drop across the simulation domain is zero due to the periodic boundary condition for fields. In the space environment, the unipolar electric fields are, however, not subject to the periodic boundary condition and hence can be viable for particle acceleration.

% figure 3

When the Landau resonant velocity is lowered further and becomes comparable to electron thermal velocity, the bipolar electric field is seen to be generated. An example with $v_r / v_{th} = 1.0$ is shown in Fig.~\ref{fig4}. Similar to the previous two cases, a beam is formed by the trapped electrons in the resonant island. However, instead of generating wave-like structures, the filament with lower phase space density in the resonant island breaks up and forms phase space holes [See Fig.~\ref{fig4}(d) and Supplementary Video 3], which correspond to the bipolar electric field in the spatial domain [Fig.~\ref{fig4}(c)]. The individual bipolar structure is approximately a few tens of Debye lengths. Rather than being regularly spaced as that of the unipolar structures, bipolar structures are more intermittent, consistent with spacecraft observations \cite{wilder2016observations}. The phase space holes are short-lived and gradually mix with other populations in the resonant island.

% figure 4

A large number of simulations \bibnote{In this parameter scan, the initial thermal velocity $v_{th}$ was varied from $0.04\, c$ to $0.01\, c$ while the Landau resonant velocity $v_r$ was fixed around $0.04\, c$. Here $c$ is the speed of light. A step size of $0.1$ for $v_r / v_{th}$ was used in the range $1.5 \leqslant v_r / v_{th} \leqslant 2.5$ to capture the transition between each regime. Outside this range, the step size for $v_r / v_{th}$ was varied from $0.2$ to $0.3$.} were performed in order to study the development and transition of the different nonlinear electric field structures, in the range $1 \leqslant v_r / v_{th} \leqslant 4$. We found that beam-mode Langmuir waves were modulated by whistler waves at approximately $2.2 \lesssim v_r / v_{th} \leqslant 4$ \bibnote{The finite amplitude of the whistler wave can only slightly affect the transition boundaries between different regimes, since the whistler parallel electric field is heavily Landau-damped and hence becomes small in the transition region.}. In this range, the whistler wave field was only slightly distorted, corresponding to weak harmonic structure in the whistler wave field. In the intermediate range $1.8 \lesssim v_r / v_{th} \lesssim 2.2$, beam-mode electron acoustic waves were generated, and the whistler wave field became highly distorted into the observed unipolar structure, simultaneously resulting in strong harmonics of whistler. In the range $1 \leqslant v_r / v_{th} \lesssim 1.8$, the bipolar electric field structure was generated. A direct comparison of $v_r / v_{th}$ between simulations and observations is difficult to perform at this stage. The cold, dense electron component of the plasma below $20$\,eV cannot be detected due to the spacecraft potential, which introduces large error bars on the measured electron thermal velocity. Furthermore, on the basis of our simple model, incorporating the observed distribution function into the simulation is also necessary for a direct comparison between simulations and observations.

We expect an amplitude threshold for whistler waves, below which the inverse distribution formed by the trapped electrons does not have a sufficiently large instability growth rate to excite the nonlinear wave structures. Determining the wave amplitude threshold from particle-in-cell simulations is not practical at this stage, since the electric field noise (due to the limited number of particles per cell) disrupts the trapping dynamics before the effect of wave amplitude threshold comes into play.

In summary, we have demonstrated that the ratio $v_r / v_{th}$ in the simulation is the controlling parameter that determines the type of electrostatic nonlinear feature that will be generated, and that the three structures observed in space in conjunction with whistler mode waves (i.e., Langmuir waves, unipolar, and bipolar structures) are all manifestations of the same nonlinear trapping phenomenon. The ratio $v_r / v_{th}$ modulates the type of emission generated by controlling the fraction of trapped electrons, the gradient of the inverse velocity distribution by the trapped electrons, and the phase velocity of associated beam-mode waves. Although the electron distributions in space are more complicated than a single Maxwellian in the simulation, our results clearly demonstrate that the trapped electrons by whistler-mode waves at different velocities of the distribution function play distinct roles in the generation of nonlinear wave structures.

\begin{acknowledgments}
This research was funded by NASA grant NNX16AG21G. We would like to acknowledge high-performance computing support from Cheyenne (doi:10.5065/D6RX99HX) provided by NCAR's Computational and Information Systems Laboratory, sponsored by the National Science Foundation. X.A. thanks G. J. Morales for fun discussions. The supplemental videos have been archived on Zenodo \url{https://doi.org/10.5281/zenodo.3959965
}.
\end{acknowledgments}

%\bibliography{chorus-nlw}

\begin{thebibliography}{38}
\bibitem [{3()}]{3}%
  \BibitemOpen
  \href@noop {} {}\bibinfo {note} {In this parameter scan, the initial thermal
  velocity $v_{th}$ was varied from $0.04\, c$ to $0.01\, c$ while the Landau
  resonant velocity $v_r$ was fixed around $0.04\, c$. Here $c$ is the speed of
  light. A step size of $0.1$ for $v_r / v_{th}$ was used in the range $1.5
  \leqslant v_r / v_{th} \leqslant 2.5$ to capture the transition between each
  regime. Outside this range, the step size for $v_r / v_{th}$ was varied from
  $0.2$ to $0.3$.}\BibitemShut {Stop}%
\bibitem [{4()}]{4}%
  \BibitemOpen
  \href@noop {} {}\bibinfo {note} {The finite amplitude of
  the whistler wave can only slightly affect the transition boundaries between
  different regimes, since the whistler parallel electric field is heavily
  Landau-damped and hence becomes small in the transition region.}\BibitemShut
  {Stop}%
\end{thebibliography}
%

%%%%%%%%%% Merge with supplemental materials %%%%%%%%%%
\pagebreak
\widetext
\begin{center}
\textbf{\large Supplementary Material to ``A unified view of nonlinear wave structures associated with whistler-mode chorus''}
\end{center}
%%%%%%%%%% Merge with supplemental materials %%%%%%%%%%
%%%%%%%%%% Prefix a "S" to all equations, figures, tables and reset the counter %%%%%%%%%%
\setcounter{equation}{0}
\setcounter{figure}{0}
\setcounter{table}{0}
\setcounter{page}{1}
\makeatletter
\renewcommand{\theequation}{S\arabic{equation}}
\renewcommand{\thefigure}{S\arabic{figure}}
%\renewcommand{\bibnumfmt}[1]{[S#1]}
%\renewcommand{\citenumfont}[1]{S#1}
%%%%%%%%%% Prefix a "S" to all equations, figures, tables and reset the counter %%%%%%%%%%

\section{The propagation direction of whistler-mode chorus}
\begin{figure}[tphb]
\centering
\includegraphics[width=6in]{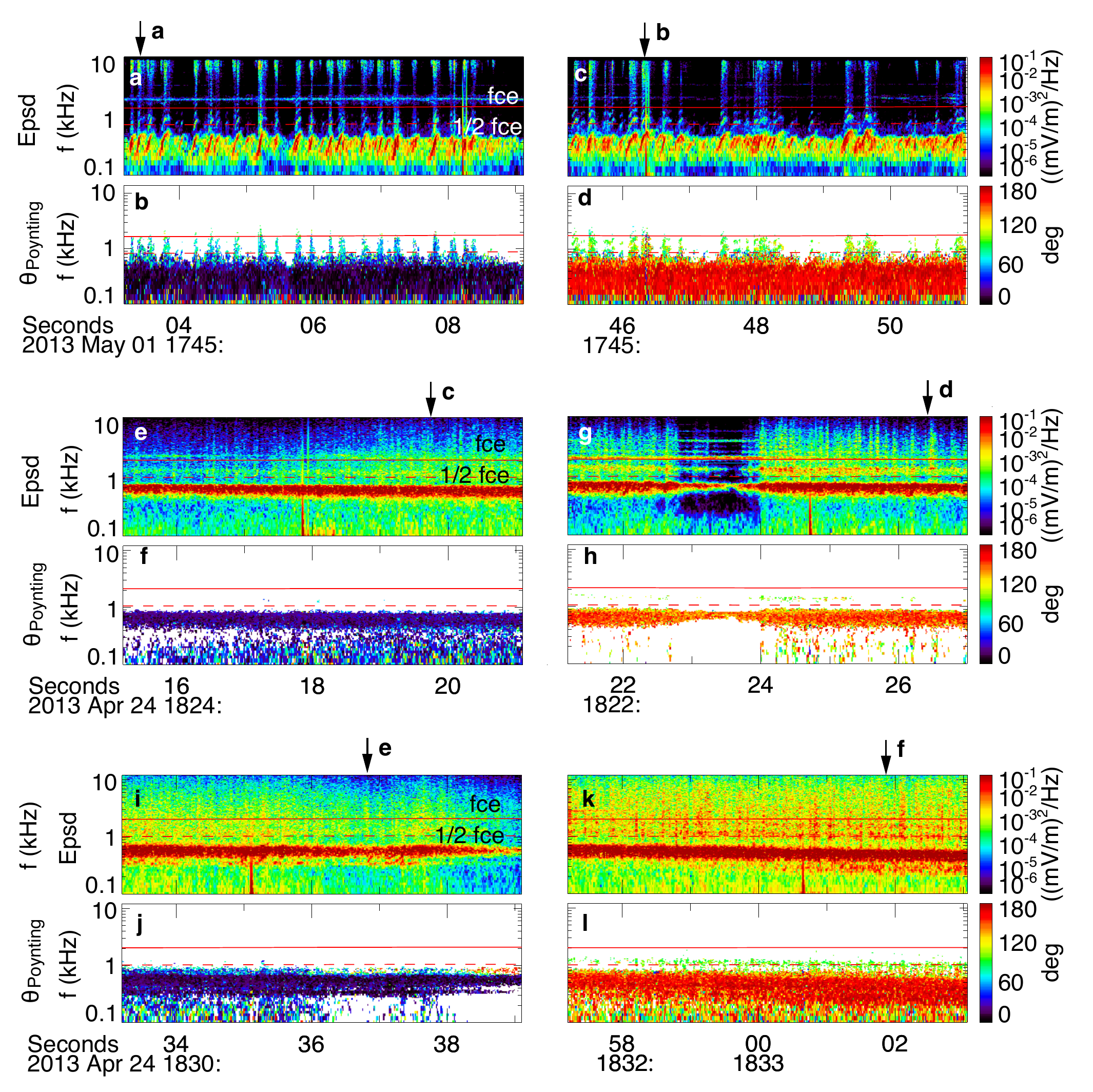}
\caption{\label{figs1}The power spectral density of electric field and the direction of the Poynting flux from spacecraft observations. (a)-(d) The spectrogram of chorus and Langmuir waves. (e)-(h) The spectrogram of chorus and its associated unipolar structures. (i)-(l) The spectrogram of chorus and its associated bipolar structures.  The direction of the Poynting flux (defined with respect to the background magnetic field) for chorus changes from parallel [panel (b), (f), (j)] to anti-parallel [panel (d), (h), (l)] in a few minutes in each example. A $30$-milliseconds time window for each case of Fig.~1 is indicated by the arrow in corresponding panel. The solid red line represents the electron cyclotron frequency whereas the dashed red line represents half of the electron cyclotron frequency.}
\end{figure}

Fig.~\ref{figs1} presents the spectrograms of electric wave power and Poynting flux direction for whistler-related Langmuir waves [Fig.~\ref{figs1}(a)-(d)], unipolar electric fields [Fig.~\ref{figs1}(e)-(h)] and bipolar electric fields [Fig.~\ref{figs1}(i)-(l)], corresponding to Van Allen Probes observation in Fig.~1. From the left panels to the right panels in Fig.~\ref{figs1}, the direction of Poynting flux changes from parallel to anti-parallel with respect to the background magnetic field. Consequently, Langmuir waves occur at the opposite chorus wave phase while unipolar and bipolar electric fields change to the opposite polarity as shown in Fig.~1. The features of spectral density are also consistent with those shown in Fig.~2(a), Fig.~3(a) and Fig.~5(a).

\section{Simulation setup}
We used a spectral Darwin particle-in-cell code in this study, which was developed as part of the University of California, Los Angeles Particle-In-Cell (UPIC) framework \cite{decyk2007upic}. The Darwin approximation neglects the transverse displacement current and hence eliminates light waves in the system, but does not affect the physics of whistler and electrostatic waves \cite{kaufman1971darwin, nielson1976particle, busnardo1977self}. The simulation has one dimension ($x$) in configuration space and three dimensions ($v_x, v_y, v_z$) in velocity space. The boundary conditions for both particles and fields are periodic. The time step $\Delta_t$ is $0.2\, \omega_{pe}^{-1}$. The cell length $\Delta_x$ is $2\, \lambda_D$, where $\lambda_D = v_{th}/\omega_{pe}$ is the initial electron Debye length. The background magnetic field $B_0$ is oriented at $30^\circ$ with respect to $x$-direction in the $x$-$z$ plane. The electron cyclotron frequency $\omega_{ce}$ is equal to $0.1\, \omega_{pe}$. This uniform background magnetic field is used since the excitation of the nonlinear wave structures is typically observed in the region around the magnetic equator \cite{li2017chorus, malaspina2018census}. The ions are treated as an immobile neutralizing background since they do not take part in the dynamics. An isotropic Maxwellian distribution is initialized for electrons. To observe the nonlinear electron trapping by whistler waves, we need to reduce the background field fluctuations to a low level compared to the chorus wave field. For this purpose, each cell contains at least $10^5$ electrons. Such computational cost is currently not affordable in two and three dimensional simulations. The detailed parameters specific for each simulation are listed in Table~\ref{table1}.

To set up the whistler wave field, we need a particle distribution that supports the wave field. To achieve this goal, we use an external pump electric field during a prescribed time interval. Therefore each electron experiences an external acceleration from the pump electric field given by
\begin{eqnarray}
\left(\frac{d v_{\alpha}}{d t}\right)_{\text{pump}} = -\frac{e}{m_e} \operatorname{Re}\left\lbrace E_{\alpha} e^{i k_0 x - i \omega_0 t} \right\rbrace,\, \alpha = x, y, z.
\end{eqnarray}
Here $v_{\alpha}$ and $E_{\alpha}$ are the particle velocity and the pump electric field, respectively. $e$ is the elementary charge, $m_e$ is the electron mass and $t$ is time. We add the pump electric field to the self-generated electric field as the total electric field, and add the background magnetic field to the self-generated magnetic field as the total magnetic field. The total electric and magnetic fields are used in the particle push. The wave number and frequency of the pump field is  $k_0$ and $\omega_0$. $k_0$ is connected to the mode number $M$ through $k_0 = 2 \pi M / (N_x \Delta_x)$, meaning the pump field has $M$ wave lengths in the system. Here $N_x$ is the number of cells in the system and $\Delta_x$ is the cell length. The associated wave magnetic field is set up naturally by the particle response. The time profile of the pump electric field is
\begin{eqnarray}
E_{\alpha} = \begin{cases}
E_{\alpha 0} \cdot (t/t_{\text{rmp}}),\, & 0 \leqslant t < t_{\text{rmp}} \\
E_{\alpha 0},\, & t_{\text{rmp}} \leqslant t < t_{\text{off}} - t_{\text{rmp}} \\
E_{\alpha 0} \cdot ((t_{\text{off}} - t)/t_{\text{rmp}}),\, & t_{\text{off}} - t_{\text{rmp}} \leqslant t < t_{\text{off}} \\
0,\, & t_{\text{off}} \leqslant t \leqslant t_{\text{end}}
\end{cases}.
\end{eqnarray}
The pump field starts with a linear up-ramp until $t = t_{\text{rmp}}$, then keeps a constant amplitude until $t = t_{\text{off}} - t_{\text{rmp}}$ and finally ends with a linear down-ramp until $t = t_{\text{off}}$. The simulation stops at $t = t_{\text{end}}$. The relative amplitude of the pump field, i.e.,  $E_{y0}/E_{x0}$ and $E_{z0}/E_{x0}$, is determined by the dispersion relation of whistler mode. The magnitude and duration of the pump field is chosen so that the normalized magnetic field $\delta B / B_0 $ of the whistler wave reaches $\lesssim 0.1$ when the pump field is turned off. Such a large amplitude wave is needed to overcome the incoherent field fluctuations in the simulation. The parameters relevant to the pump field are listed in Table~\ref{table1}.

Finally, it is worthy to note that the setup of the problem in the simulation is a temporal problem whereas it is a spatial problem in the space environment. In other words, the nonlinear trapping occurs across the simulation domain but for a limited time span in simulations, while in the space environment it occurs in a limited spatial range but for a longer time span when the whistler waves propagate from lower to higher latitudes. Nevertheless, the underlying physics of nonlinear trapping is the same in both scenarios.

\begin{table}
\centering
\begin{tabular}{l *{6}{c}}
 \hline
               & $N_x$ & $\Delta_x \omega_{pe} /c$   & $v_{th}/\Delta_x \omega_{pe}$ & $M$ & $\omega_0/\omega_{pe}$ & $v_r / v_{th}$ \\
 \hline
Simulation 1   & $4096$ & $0.0267$  & $0.5$   & $10$ & $0.0215$ & $3.2$ \\
Simulation 2   & $2048$ & $0.04$    & $0.5$     & $7$  & $0.0194$ & $2.1$ \\
Simulation 3   & $1024$ & $0.08$    & $0.5$     & $7$  & $0.0194$ & $1.0$ \\
 \hline
               & $\tilde{E}_{x0}$ & $\tilde{E}_{y0}$ & $\tilde{E}_{z0}$ & $t_{\text{rmp}}\omega_{pe}$ & $t_{\text{off}}\omega_{pe}$ & $t_{\text{end}}\omega_{pe}$ \\
 \hline
Simulation 1   & $1.39$ & $1.81 i$ & $-1.81$ & $62.83$ & $1100$ & $4000$ \\
Simulation 2   & $1.35$ & $1.81 i$ & $-1.81$ & $62.83$ & $750$  & $4000$ \\
Simulation 3   & $1.35$ & $1.81 i$ & $-1.81$ & $62.83$ & $500$  & $4000$ \\
 \hline
\end{tabular}
\caption{\label{table1}The parameters for three nominal simulations. The electric field is normalized as $\tilde{E}_{\alpha 0} = 10^4 e E_{\alpha 0} / (m_e \omega_{pe}^2 \Delta_x)$ with $\alpha = x, y, z$. The imaginary unit $i$ in the value of $E_{y0}$ represents a phase shift of $90^\circ$. The negative sign in the value of $E_{z0}$ represents a phase shift of $180^\circ$.}
\end{table}

\end{document}